\documentclass[letterpaper]{article} 
\usepackage{aaai25}  
\usepackage{times}  
\usepackage{helvet}  
\usepackage{courier}  
\usepackage[hyphens]{url}  
\usepackage{graphicx} 
\usepackage{amsmath}
\usepackage{natbib}  
\usepackage{caption} 
\usepackage{algorithm}
\usepackage{algorithmic}
\usepackage[table]{xcolor}
\usepackage{subcaption}
\usepackage{placeins} 

\usepackage{newfloat}
\usepackage{listings}
\DeclareCaptionStyle{ruled}{labelfont=normalfont,labelsep=colon,strut=off} 
\floatstyle{ruled}
\newfloat{listing}{tb}{lst}{}
\floatname{listing}{Listing}
\title{Unmasking Superspreaders: Data-Driven Approaches for Identifying and Comparing Key Influencers of Conspiracy Theories on X.com}
\author {
    Florian Kramer\textsuperscript{\rm 1},
    Henrich R. Greve\textsuperscript{\rm 2},
    Moritz von Zahn\textsuperscript{\rm 3},
    Hayagreeva Rao\textsuperscript{\rm 1}
}
\affiliations {
    \textsuperscript{\rm 1}Graduate School of Business, Stanford University, 655 Knight Way, Stanford, CA 94305, USA\\
    \textsuperscript{\rm 2}Entrepreneurship Area, INSEAD, 1 Ayer Rajah Ave, Singapore 138676, Singapore\\
    \textsuperscript{\rm 3}Information Systems Department, Goethe University, Theodor-W.-Adorno-Platz 1, 60323 Frankfurt am Main, Germany\\
     florian.kramer@ok.de, henrich.greve@insead.edu, vzahn@wiwi.uni-frankfurt.de, hrao@stanford.edu
}

\usepackage{bibentry}
\nocopyright

\begin{document}

\maketitle

\begin{abstract}
Conspiracy theories can threaten society by spreading misinformation, deepening polarization, and eroding trust in democratic institutions. Social media often fuels the spread of conspiracies, primarily driven by two key actors: Superspreaders---influential individuals disseminating conspiracy content at disproportionately high rates, and Bots---automated accounts designed to amplify conspiracies strategically. To counter the spread of conspiracy theories, it is critical to both identify these actors and to better understand their behavior. However, a systematic analysis of these actors as well as real-world-applicable identification methods are still lacking. In this study, we leverage over seven million tweets from the COVID-19 pandemic to analyze key differences between Human Superspreaders and Bots across dimensions such as linguistic complexity, toxicity, and hashtag usage. Our analysis reveals distinct communication strategies: Superspreaders tend to use more complex language and substantive content while relying less on structural elements like hashtags and emojis, likely to enhance credibility and authority. By contrast, Bots favor simpler language and strategic cross-usage of hashtags, likely to increase accessibility, facilitate infiltration into trending discussions, and amplify reach. To counter both Human Superspreaders and Bots, we propose and evaluate 27 novel metrics for quantifying the severity of conspiracy theory spread. Our findings highlight the effectiveness of an adapted H-Index for computationally feasible identification of Human Superspreaders. By identifying behavioral patterns unique to Human Superspreaders and Bots as well as providing suitable identification methods, this study provides a foundation for mitigation strategies, including platform moderation policies, temporary and permanent account suspensions, and public awareness campaigns.
\end{abstract}

%

\section{Introduction}
In the face of high uncertainty or perceived threats, humans often make sense of complex events by turning to conspiracy theories. A major reason is that conspiracy theories satisfy the human need for causal explanations, making complex events feel more understandable \citep{PyschoOfConspiraDouglas}. This is especially the case for major incidents like the  January 6th Capitol attack or the COVID-19 pandemic, where conspiracy theories are particularly reinforced by sparse or ambiguous information \citep{DramaticEventsSamory}. Conspiracy theories offer alternative, typically unverified explanations, and their self-sealing nature, where attempts to disprove them are co-opted as further evidence, further reinforce their credibility among adherents \citep{ConspiracySunstein}. Conspiracy theories are widespread, with nearly half of the U.S. population believing in at least one \citep{ConspiracyOliver}.

Conspiracy theories can have harmful consequences and even cost lives. Specifically, these theories can erode trust in key institutions such as governments and scientific bodies, diminish political participation, and deepen social polarization \citep{ConsequencesConspiracyTaylor}. For example, during the COVID-19 pandemic, they weakened public health initiatives by undermining scientific guidance and promoting behaviors that contradicted established health guidelines \citep{ConspiracyCovidDow}. 

Conspiracy theories emerge and spread particularly strongly through users on social media. The theories often originate within online groups, where users share interpretations and seek validation from like-minded individuals \citep{VaccineGrant}. The algorithms employed by these platforms often amplify misleading content, and research indicates that false information travels farther and faster than verified news \citep{TheSpreadOfTrueAndFalseVosoughi}. To curb the spread of conspiracy content, platforms often employ targeted countermeasures, including content removal and temporary or permanent user suspensions \citep{innes2023platforming}.

Two types of users are central to the spread of conspiracy theories on social media, namely ``Human Superspreaders''---highly influential accounts of human users that disseminate disproportionate amounts of conspiracy content---and ``Bot Spreaders''---automated accounts programmed to amplify such content through strategic tactics, such as targeting susceptible users and exploiting psychological tendencies. While both play critical roles, Human Superspreaders and Bot Spreaders differ fundamentally in nature and behavior. Authors such as \citet{IdentifySuperspreaderDeVerna, FakeNewsGrinberg, DisinformationDozenNogara, SuperspreaderBodaghi} emphasize the substantial impact of Human Superspreaders on the spread of conspiracy content.\footnote{Since we are the first to contrast human and bot accounts in spreading misinformation, prior work does not explicitly distinguish between the two. However, given that prior findings often center on a small set of highly influential accounts known to be real people, it is reasonable to assume that a substantial share of the problem stems from Human Superspreaders.} For example, \citet{FakeNewsGrinberg} found that during the 2016 U.S. election, just 0.1\% of Twitter accounts generated 80\% of fake news, accounting for approximately 6\% of total news consumption. Similarly, twelve individuals, dubbed the ``Disinformation Dozen,'' were responsible for 65\% of COVID-19 misinformation \citep{center2021disinformationdozen}. Conversely, other research highlights the significant role of automated bot accounts \citep{OnlineConspiracyGreve, BotInteractionVarol, LowCredContentShao, TwitterFacebookYang}. For example \citet{LowCredContentShao} demonstrates that they play a disproportionate role in the spread of untrustworthy content by strategically targeting high-profile users through replies and mentions to boost content visibility.

Both Human Superspreaders and Bot Spreaders contribute to the dissemination of conspiracy content, yet distinguishing between them is crucial for designing effective countermeasures. Since bot accounts are not real humans, they lack rights to free expression and, arguably, do not directly contribute to platform revenue. Platforms can therefore suspend such accounts permanently with no unintended side effects. Human Superspreaders, by contrast, are individuals with rights to freedom of expression, and restrictions on their accounts must be carefully weighed against the benefits of reducing misinformation. Moreover, interventions against Human Superspreaders may generate unintended consequences, as documented in research on “de-platforming” \citep[see, e.g.,][]{innes2023platforming}. It is thus essential to distinguish between and better understand the characteristics of conspiracy content spread by Human Superspreaders and Bot Spreaders.

To the best of our knowledge, there is no established method for identifying Human Superspreaders and distinguishing them from Bot Spreaders. Moreover, despite the central role of both actors and the importance of differentiating between them, no comparative studies to date have systematically examined these two groups in relation to other user types on social media platforms. This study aims to contribute to closing this important gap by addressing the following research questions: \\ \textbf{How do Human Superspreaders and Bot Spreaders differ from other users in their behavior on social media? How can we effectively identify them using public data only?} 

We address these research questions by developing a practical and computationally feasible approach to identify these actors and by systematically analyzing their behavior. Specifically, we first provide a suitable metric that enables the detection and suspension of accounts in a way that curbs the spread of conspiracy content efficiently, that is, achieving the largest reduction in spread with a limited number of suspensions. For this, we propose and evaluate 27 candidate metrics to rank users spreading misinformation according to their severity of influence. We rely on a dataset of over seven million COVID-19-era conspiracy-related tweets \citep{OnlineConspiracyGreve} in which we identify conspiracy content using a prompt-tuned BLOOMZ-1.7B (BigScience Large Open-science Open-access Multilingual Language Model) auto-regressive LLM (Large Language Model) \citep{PaulRabbithole}. We then use Botometer \citep{botometerX} to detect Bot Spreaders and then compare linguistic, behavioral, emotional, and political characteristics of Human Superspreaders, Bot Spreaders, Human Spreaders (occasionally spreading conspiracy content), and Human Non-Spreaders (never spreading such content). Our findings illuminate how these groups differ in their content and roles, thereby contributing to a more nuanced understanding of their respective roles in spreading misinformation and providing guidance for platform owners in their response to both actor types.

\section{Related Work}

Current methods for identifying Superspreaders (human or automated) rely heavily on computationally intensive techniques such as machine learning and natural language processing, limiting their accessibility and adaptability \citep{AutomaticDetectionSuperspreaderSmith}. Structural approaches like k-core decomposition and Topic-Sensitive PageRank often make unrealistic assumptions about network dynamics \citep{KCoreAlvarez, TopicPageRankHaveliwala}, while simpler metrics, such as the H-Index adaptation by \citet{IdentifySuperspreaderDeVerna} which defines Human Superspreaders as accounts with at least h tweets and h retweets demonstrate computational efficiency, but primarily focus on retweets and overlook other engagement forms like replies, likes, and quotes. To address these limitations, this study proposes incorporating a broader range of engagement parameters, balancing computational efficiency with greater contextual richness. 

Beyond the mere identification of actors spreading conspiracy content, a deeper understanding of \emph{how} they disseminate it remains crucial yet largely lacking.
In fact, the existing approaches for identification typically operate as black boxes, offering limited interpretability despite their potentially high impact when used to suspend human accounts, a practice that scholars caution against and argue should be based on interpretable, understandable methods \citep[see, e.g.,][]{rudin2019stop}. Current work therefore fail to provide insights into the distinct and overlapping behaviors of Human Superspreaders and Bot Spreaders. As our analysis shows, the former typically generate nuanced, context-specific content that requires more sophisticated interventions, whereas the latter produce repetitive, structured misinformation that is more amenable to automated detection. Understanding these differences is essential for developing targeted and interpretable strategies to mitigate the spread of conspiracy theories, for example, tailored psychological interventions aiming to increase people's ability to discern trustworthy from untrustworthy content \citep[see, e.g.,][]{roozenbeek2022psychological}.

\section{Data}

\textit{Overview.} We analyze 7,616,569 tweets from 7,416 unique users collected by \citet{OnlineConspiracyGreve} during the COVID-19 pandemic. Since the pandemic emerged in December 2019, online communities have been rife with speculation on the virus’ origin, impact, and very existence \citep{SociologyConspiracyRao}. These conspiracy theories undermine public health efforts by eroding trust in experts, promoting mask resistance, and discouraging vaccination \citep{CovidNotWearingMaskPrichard, ConspiracyConsequencesCovidRomer}. Their prevalence was significant enough for U.S. President Joe Biden to accuse social media platforms of allowing content that harmed vaccination rates and “killed people” \citep{biden_facebook_2021}. Despite this urgency, reliably quantifying who spreads such narratives has been challenging. The dataset from \citet{OnlineConspiracyGreve} offers a valuable opportunity to address these gaps.

\textit{Data Collection.} \citet{OnlineConspiracyGreve} used Twitter’s formerly publicly accessible API (before its rebranding to X) to collect the data in a two-stage process. First, they collected all tweets containing the hashtags \#coronahoax and \#virushoax posted between February 17 and May 22, 2020. Due to API rate limits, focusing on these specific hashtags ensured a manageable query volume \citep{StreamingTwitterCampan}. Second, they retrieved the complete tweet history (from January 2020 onward) of users who had used these hashtags.

\textit{Data Preprocessing.} To allow a broader search for influential accounts, we focus on retweeted content. 
If any of user retweeted another account (e.g., that of Donald Trump), we included the original tweet to assess the influence of the ``Original Tweeter'' whereas the user who performed the retweet is referred to as the “Retweeter”. From the 7,616,569 collected tweets, only retweets were retained as they are generally interpreted as endorsements \citep{IdentifySuperspreaderDeVerna}, leaving 2,791,823 retweets. Next, to ensure the generation of user-level metrics, we excluded retweets for which the corresponding user data could not be obtained
, yielding 2,349,333 retweets. We refined the dataset by filtering tweets to include only those from accounts actively endorsing conspiracy theories. Using an ensemble of prompt-tuned BLOOMZ-1.7B (BigScience Large Open-science Open-access Multilingual Language Model), an auto-regressive large language model (LLM) developed by \citet{PaulRabbithole}, we classified tweets as endorsing conspiracy theories if the model assigned a positive classification probability exceeding 90\%. Retweets from users with at least one conspiracy-endorsing post were retained, resulting in 954,993 retweets from 7,043 users. Finally, duplicate retweets and records missing engagement data (like counts or retweet counts) were removed. The resulting dataset for our analysis comprises 261,371 retweets from 7,043 Original Tweeters.

\section{Methodology}

To analyze Human Superspreaders and Bot Spreaders engaged in spreading conspiracy theories, we (1)~propose metrics to classify and rank all Superspreaders, (2)~conduct a dismantling analysis to evaluate these metrics, and (3)~perform extensive content analyses to uncover behavioral and linguistic patterns across Human Superspreaders, Bot Spreaders, Human Spreaders, and Human Non-Spreaders.

\subsection{Metrics to Classify and Rank Superspreaders}

\textit{Human Superspreaders.} To systematically evaluate the influence of Human Superspreaders, we propose a novel set of metrics. Importantly, these metrics build upon a broader set of engagement-based inputs (e.g., Likes, Quotes, Replies) that, to the best of our knowledge, have not previously been incorporated in this context. Whereas prior research has typically assessed the influence of Human Superspreaders based primarily on the number of followers or retweets \citep[see, e.g.,][]{IdentifySuperspreaderDeVerna}, our approach integrates additional forms of user interaction to capture a more comprehensive picture of influence. Specifically, in our study, we include Replies, Likes, and Quotes and two composite metrics that we term Engagement Score \((ES)\) and Normalized Engagement Score (\(ES^\mathrm{norm}\)). 

Formally, the Engagement Score of the \(i\)-th tweet (\(ES_i\)) is defined as 

\begin{equation}
ES_i = RT_i + R_i + L_i + Q_i,
\end{equation} 

where \(RT_i\), \(R_i\), \(L_i\), and \(Q_i\) represent the counts of Retweets, Replies, Likes, and Quotes, respectively. Accordingly, the Normalized Engagement Score is calculated by normalizing each Engagement: 

\begin{equation}
ES_i^{\text{norm}} = RT_i^{\text{norm}} + R_i^{\text{norm}} + L_i^{\text{norm}} + Q_i^{\text{norm}}
\end{equation} 

where \(X_i^{\text{norm}} = \frac{X_i - \min(X)}{\max(X) - \min(X)}\) for each Engagement \(X\).
Using these Engagements, we compute 27 metrics organized into five categories:
\begin{enumerate}

    \item \textit{Aggregate Metrics}:  
    Sum of each Engagement \(X\) across all tweets by the same user, that is, 
    \begin{equation}
    \label{eq:aggr}
        \text{Aggregate Metric}_u(x) = \sum_{i=1}^{N_u} x_i,
    \end{equation}
    where \(x_i\) represents a specific Engagement (e.g., Likes) for the \(i\)-th tweet, and \(N_u\) is the total number of tweets authored by user \(u\).

    \item \textit{Per-Tweet Metrics}:  
    Average engagement per tweet to highlight users who achieve consistent traction with fewer tweets, that is, 
    \begin{equation}
        \text{Per-Tweet Metric}_u(x) = \frac{\text{Aggregate Metric}_u(x)}{N_u}.
    \end{equation}
    
    \item \textit{Follower-Weighted Aggregate Metrics}:  
    Tweet’s engagement weighted by the follower count of the most prominent Retweeter, which captures the amplified reach conferred by high-follower accounts and is formally defined as
    \begin{equation}
        \text{Follower-Weighted Aggr. Metr.}_u(x) = \sum_{i=1}^{N_u} x_i \cdot f_i, 
    \end{equation}
    where \(f_i\) is the number of followers of the Retweeter with the highest follower count for the \(i\)-th tweet.
    
    \item \textit{Follower-Weighted Per-Tweet Metrics}:  
    Average follower-weighted engagement per tweet, that is,
    \begin{equation}
        \text{Foll.-W. Per-Tw. Metr.}_u(x) = \frac{\text{Foll.-W. Aggr. Metr.}_u(x)}{N_u}.
    \end{equation}
    
    \item \textit{Adapted Scholarly Metrics}: 
    To measure the consistent influence of a user, \citet{CovidGallagher} propose a modified version of the \textit{H-Index} \citep{HIndexHirsch}, originally designed to assess researcher impact. 
    In Gallagher’s adaptation, the \textit{H-Index} \( h(i) \) is defined as the maximum value of \( h \) for which user \( i \) has authored at least \( h \) tweets, each retweeted at least \( h \) times. An account achieves a high \textit{H-Index} by authoring multiple conspiracy tweets that are widely retweeted. Users with a single viral tweet or low-engagement tweets will have a low \textit{H-Index}. For example, a user \(i\) with \(h = 10\) has at least 10 tweets, each retweeted 10 times. Formally, 
    \begin{equation}
        h(i) = \{ h \mid h \text{ tweets have at least } h  \text{ retweets each}\}    
    \end{equation}
    
Building on \citet{CovidGallagher}, we propose the \textit{M-Index}, which adjusts the \textit{H-Index} for the time period of activity to reflect a user’s sustained rather than cumulative influence: 
\begin{equation}
    M(i) = \frac{h(i)}{T(i)}, 
\end{equation}
where \( T(i) \) is the number of months since user \( i \)'s first tweet in the dataset.
    
Finally, we assess the \textit{G-Index} \citep{GIndexEgghe}, which assigns greater weight to highly retweeted tweets, addressing limitations of the \textit{H-Index} \citep{HIndexGIndexManjareeka}: 
\begin{equation}
    \label{eq:gindex}
    g(i) = \max \left\{ g \mid \sum_{j=1}^{g} RT_j \geq g^2 \right\},     
\end{equation}
where \(RT_j\) are the retweet counts of the top \(g\) tweets authored by user \(i\).

\end{enumerate}

\subsection{Dismantling Analysis to Evaluate Metrics}

To evaluate the 27 metrics we follow \citet{LowCredContentShao} and apply a dismantling analysis. This involves sequentially removing Original Tweeters and their tweets from the Retweet Dataset in ascending order of their rank according to each metric. The reduction in total retweet volume is then analyzed. Metrics are deemed more effective if removing a small number of users results in a significant drop in retweet volume, indicating the critical role of those users in spreading conspiracy content. Formally, the impact of removing an Original Tweeter \( i \) is defined as 

\begin{equation}
    \text{Impact}_i = \frac{\text{Number of tweets removed for original user } i}{\text{Total tweets in the Retweet Data Set}},     
\end{equation}

where the numerator represents the tweets eliminated by removing user \( i \), and the denominator is the total number of conspiracy-related tweets in the dataset.

To benchmark the metrics, we introduce an Optimal Metric as a theoretical baseline. This metric assumes perfect knowledge of the Twitter Network and ranks users in descending order of their total contribution to conspiracy-related tweets. While this idealized metric is infeasible in real-world settings due to the complexity and dynamic nature of social networks, it provides a useful reference for comparing the relative effectiveness of practical metrics.

\subsection{Identification of Bot Spreaders} We identify Bot Spreaders using Botometer X \citep{botometerX}, the latest version of Botometer, a prominent machine-learning classifier for Bots used in hundreds of scientific papers developed by the Observatory on Social Media (OSoMe) at Indiana University \citep{OnlineConspiracyGreve}. Botometer X evaluates over 1,000 features across six categories—user profile, friends, network structure, temporal patterns, content and language use, and sentiment analysis—to calculate a conditional probability that an account is automated \citep{BotInteractionVarol}. Notably, the linguistic features of Bot Spreaders that we later analyze may overlap with features considered by Botometer. In principle, this could introduce endogeneity concerns: observed linguistic differences between Bot Spreaders and humans might partly reflect the same language cues Botometer uses for classification. However, note that Botometer evaluates a far broader and more diverse set of features, including network structure, temporal metadata, and interaction patterns. We therefore follow prior work that relies on Botometer as a well-validated foundation for bot identification and proceed to analyze the dissemination patterns of the detected bots \citep[e.g.,][]{{BotsCovidTwitterFerrara}}. 

\subsection{Comparative Content Analyses}

Human Superspreaders and Bot Spreaders are pivotal drivers of misinformation on social media, utilizing distinct strategies to amplify conspiracy theories. To develop effective mitigation strategies, we differ between and compare these groups, as well as with Human Spreaders (users who share conspiracy content less frequently) and Human Non-Spreaders (those who do not share conspiracy content). Our analysis examines various linguistic and behavioral features, enabling us to identify how these groups differ in terms of content, sentiment, emotions, toxicity, and political orientation.

\textit{Content Analysis.} We conduct a content analysis to uncover linguistic patterns and variations in shared material.

\begin{enumerate}
    \item \textit{Language Complexity and Readability}:
    To examine differences in tweet readability, defined by \citet{klare1963measurement} as ``the ease of understanding or comprehension due to the style of writing''
    , we follow \citet{FleschKincaidTwitterAhmed} and apply the widely recognized Flesch–Kincaid formula. This method estimates the U.S. grade level required to understand a text based on average sentence length and syllables per word \citep{ReadabilityFleschKincaidKayam}. Following \citet{VaderShihab}, we preprocess the tweets by removing non-informative elements such as URLs, Retweet-Tags, and mentions, while retaining hashtags for contextual purposes and performing a minor outlier adjustment
    \footnote{0.23\% of tweets that were nonsensical has excessive Flesch-Kincaid scores above 25 and were removed.}.

    \item \textit{Linguistic Features and Content Characteristics}:
    To investigate content differences, we analyze key tweet features using binary variables (1 if present, 0 otherwise) as recommended by \citet{HateSpeechTwitterMaarouf}. The analyzed features include mentions, hashtags, media items, emojis, exclamation marks, question marks, and all-caps usage. These were selected for their relevance in signaling and shaping communication:

    \begin{itemize}
        \item \textit{Mentions} (e.g., “@username”) often target specific individuals, contrasting with generalized content, which spreads differently \citep{HateSpeechMai}.
        \item \textit{Hashtags} (e.g., “\#hashtag”) act as thematic markers and visibility tools \citep{HateSpeechTwitterMaarouf}.
    	\item \textit{Media Items} (e.g., images, videos, polls, URLs) enhance engagement and reach \citep{HateSpeechTwitterMaarouf}.
        \item \textit{Emojis} are communication signals within conspiracy networks \citep{EmojiTwitterQAnonWang, EmojiConspiracyGualda}.
        \item \textit{Exclamation marks} amplify intensity, boosting engagement for positive comments and enhancing negativity in critical ones \citep{ExclamationMarksTeh}. \textit{Question marks} are analyzed for comparable effects.
        \item \textit{All-caps text} conveys emotions such as excitement, emphasis, and urgency, influencing the emotional tone of tweets \citep{AllCapsHeath}.
    \end{itemize}
    
    In line with \citet{HateSpeechTwitterMaarouf}, we also measure the raw length of tweets (excluding URLs, mentions, and hashtags) alongside their unedited length to capture differences in overall feature usage.

    \item \textit{Hashtags and Co-occurrence Networks}:
    Hashtags play a critical role in content dissemination, increasing visibility and fostering subcommunities by associating post with relevant topics and connecting like-minded users \citep{HashtagChakrabarti, HashtagCommunityLledo}. In conspiracy theories, hashtags serve as thematic markers and tools for promoting narratives or challenging opposing views \citep{HashtagsPoliticalDirectionStewart}. In addition, pairing popular hashtags with niche ones can amplify lesser-known topics, such as using hashtags associated with Donald Trump to promote unrelated conspiracies \citep{CoHashtagMonaci}.
    
    To analyze strategic hashtag usage, we rank the most frequently used hashtags by user group and construct co-hashtag networks. In these networks, hashtags are nodes, edges represent co-occurrence, and edge thickness reflects frequency. Node thickness indicates overall usage, revealing patterns in hashtag strategies and visibility enhancement \citep{CoHashtagMonaci}.

    \item \textit{Word Frequency}: We further compare groups by following authors like \citep{WordCloudFong} by counting the most frequently used words and visualizing prominent terms using word clouds, in which word size represents frequency \citep{WordCloudFong, WordCloudSkeppstedt}.  Before analysis, we remove stop words using the WordCloud class in Python, which excludes 192 common filler words such as "an," "and," "the," and "they" \citep{amueller_wordcloud_stopwords}.
\end{enumerate}

\textit{Sentiment, Emotional, and Toxicity Analysis.} Utilizing natural language processing, we next identify and evaluate opinions, sentiments, attitudes, and emotions expressed in tweets \citep{VaderGilbert2014}. Following \citet{VaderShihab}, we once again preprocess the tweets by removing non-informative elements such as URLs, Retweet-Tags, and mentions, while retaining hashtags for contextual purposes. 

\begin{enumerate}
    \item \textit{VADER Sentiment}: We employ NLTK's VADER (Valence Aware Dictionary for Sentiment Reasoner) analyzer, optimized for microblog contexts, to classify tweets as positive, negative, or neutral based on compound scores \citep{VaderGilbert2014}. Each word in VADER is assigned a sentiment rating and intensity, indicating its positivity, negativity, or neutrality \citep{SentimentAnalysisTwitterShelar}. Following \citet{VaderShihab}, tweets with compound scores $> 0.001$ are labeled positive, between -0.001 and 0.001 as neutral, and $< -0.001$ as negative.

    \item \textit{Ekman’s Basic Emotions}: To identify the emotions expressed in a tweet, we follow the approach of \citet{HartmannEmotionsTweetsButt} and utilize a fine tuned checkpoint of the DistilRoBERTa base model developed by \citet{BERTSanh2019}. This model is trained on a combination of multiple emotion datasets for English and predicts Ekman’s six basic emotions (anger, disgust, fear, joy, sadness, and surprise), along with a neutral class \citep{hartmann2022emotionenglish}.

    \item \textit{Tweet Toxicity}: Following the approach of authors like \citet{ToxicMosleh2022} and \citet{IdentifySuperspreaderDeVerna}, we utilize the Google Jigsaw Perspective API, to assess the toxicity of tweets. The API characterizes toxic language as remarks that are rude, disrespectful, or unreasonable, potentially discouraging users from participating in online discussions \citep{IdentifySuperspreaderDeVerna, PerspectiveAPI}. To do so it utilizes machine learning models to detect abusive language, assigning a score that reflects the potential negative impact of the text on a conversation \citep{PerspectiveAPI}. Despite the model’s "black box" nature, it aligns with established methodologies in social media research \citep{IdentifySuperspreaderDeVerna}.

\end{enumerate}

\textit{Political Orientation Analysis.} Building on the work of \citet{HashtagsPoliticalDirectionStewart} and \citet{HashtagsLeftRightStewart}, we utilize hashtags to determine the political orientation of users. Accodingly, we assess political orientation by manually classifying the 500 most frequent hashtags as left-leaning, right-leaning, or non-political, resolving disagreements through consensus among three researchers. Tweets and accounts are then labeled according to the predominant orientation of their hashtags, while ties or ambiguous cases are classified as non-political.



\section{Results}

In the following, we first present the dismantling analysis of our proposed metrics, followed by comparative content analyses of Human Superspreaders and Bot Spreaders.

\subsection{Dismantling Analysis}

\begin{figure}[!h]
    \centering
    \includegraphics[width=1\linewidth]{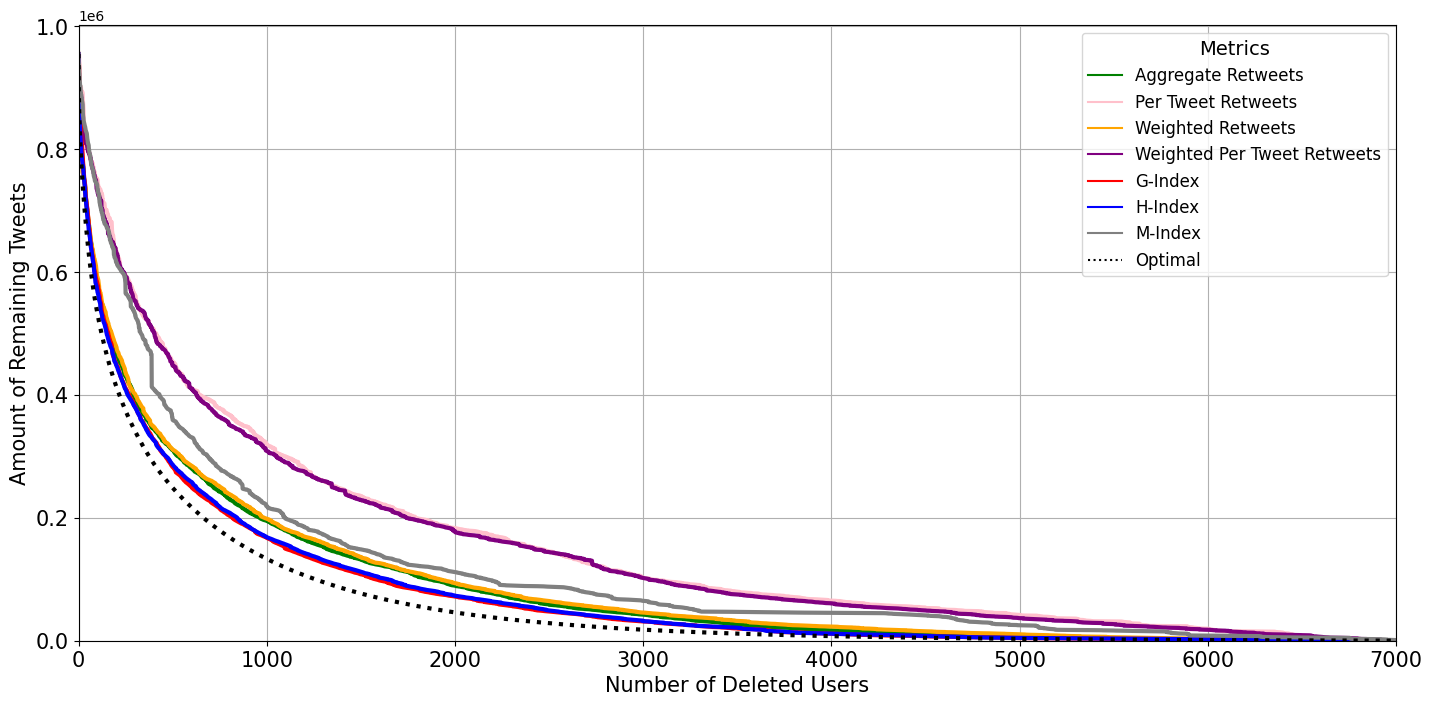}
    \caption{Dismantling Analysis for Identifying the Most Influential Human Superspreaders}
    \label{fig:Dismantling Analysis}
\end{figure}

\begin{figure}[!h]
    \centering
    \includegraphics[width=1\linewidth]{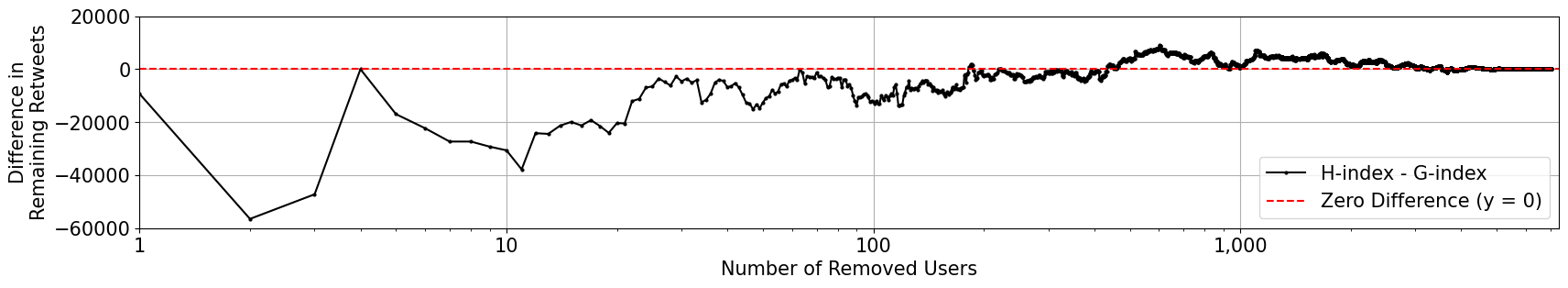}
    \caption{Comparison of H-Index and G-Index using Difference in Remaining Users}
    \label{fig:H-Index vs. G-Index: Difference in Remaining Users}
\end{figure}

To identify the most influential Human Superspreaders, we ranked users via the 27 metrics proposed and performed a dismantling analysis \citep{LowCredContentShao}. Figure~\ref{fig:Dismantling Analysis} depicts the remaining number of conspiracy-related tweets (vertical axis) as a function of the number of users removed based on their ranking under each approach (horizontal axis) based on retweets.\footnote{Figure~\ref{fig:Dismantling Analysis} reports results for retweets only, as other forms of engagement (e.g., Likes and Replies) yield comparable performance within each approach.} Accordingly, steep drops in tweet volume upon removing few users indicate stronger performance of the proposed approaches. Overall, Per-Tweet metrics performed worst, while Aggregate and Follower-Weighted metrics showed moderate success. Two \textit{Adapted Scholarly Metrics}—the \textit{H-index} and \textit{G-index}—outperformed other approaches, with the \textit{M-index} lagging behind. As shown in Figure~\ref{fig:H-Index vs. G-Index: Difference in Remaining Users}, the H-index performs slightly better than the G-index when only a small number of users are removed (i.e., few accounts suspended), whereas the G-index proves more effective once larger numbers of users are removed.

\begin{figure}[!h]
    \centering
    \includegraphics[width=1\linewidth]{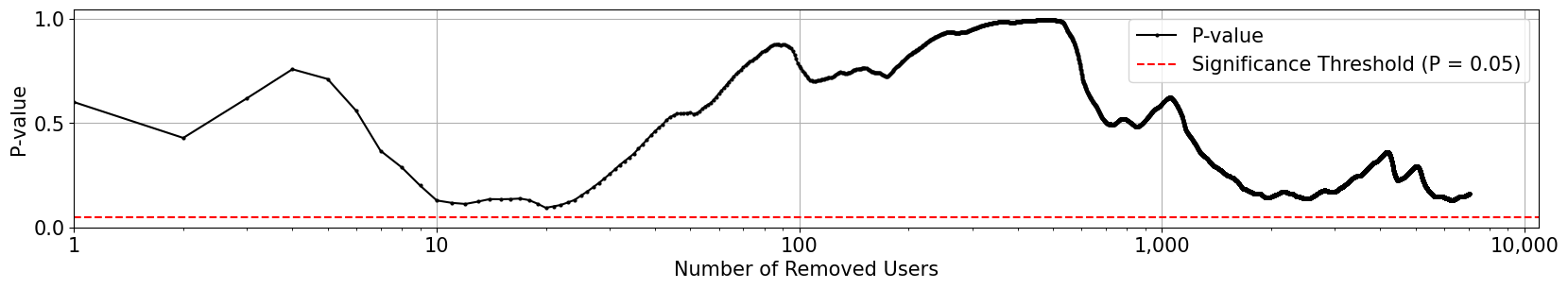}
    \caption{Comparison of H-Index and G-Index using Cramér–von Mises Two-Sample Test}
    \label{fig:H-Index vs. G-Index: Cramér–von Mises two-sample test}
\end{figure}

If platforms aim to de-platform only Superspreaders, that is, to permanently suspend accounts only in extreme cases of misinformation, the performance at very low suspension levels is most relevant. For instance, the removal of just 0.1\% of users — the threshold we selected for classifying superspreaders — based on the H-index resulted in the elimination of seven human superspreaders, who collectively accounted for 12.35\% of all conspiracy-related tweets in our dataset. This reduction comes close to the 13.43\% achieved by the optimal benchmark, underscoring the suitability of this metrics for identifying Human Superspreaders in this range. However, both the H-index and G-index appear effective at such low levels, as the Cramér–von Mises two-sample test in Figure~\ref{fig:H-Index vs. G-Index: Cramér–von Mises two-sample test} shows no statistically significant difference between the two.

\subsection{Comparative Analysis of Human Superspreaders to Bot Spreaders}

\textit{Content Analysis.} We conduct a content analysis to uncover linguistic patterns and variations in conspiracy-related tweets between Human Superspreaders and Bot Spreaders. Importantly, following \citet{BotOrNotDavis}, we classify an account as a Bot if its Botometer score exceeds 0.4, revealing 918 out of 7,043 users as Bot Spreaders (13.03\% of total users) that are responsible for 35,109 conspiracy-related tweets.

\begin{figure}[!h]
    \centering
    \includegraphics[width=1\linewidth]{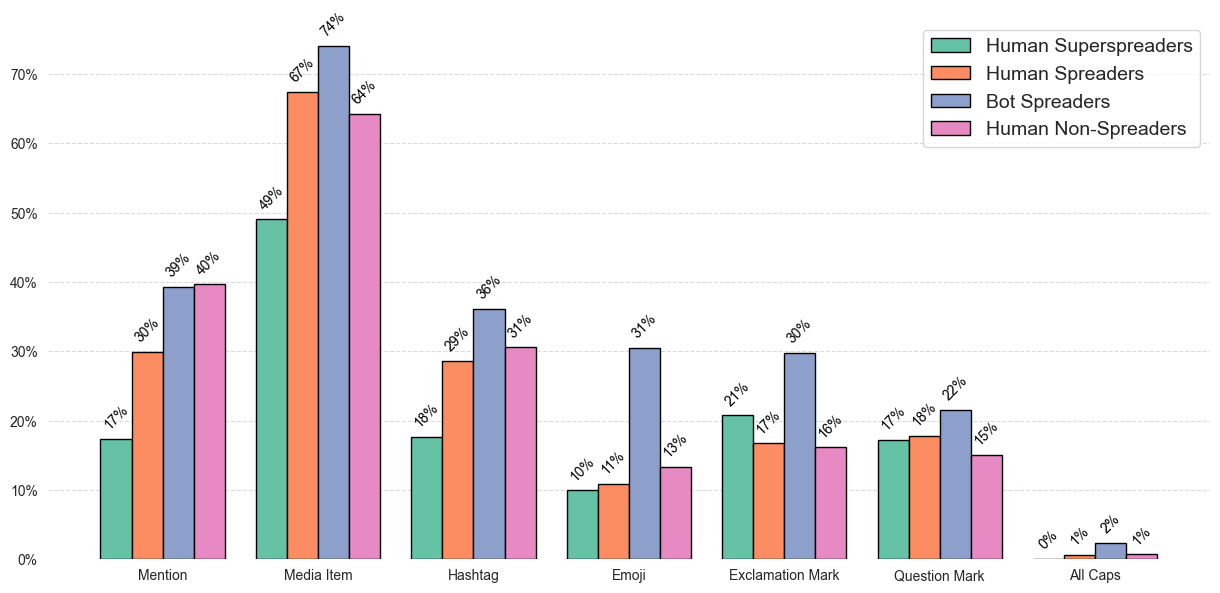}
    \caption{Analysis of Linguistic Features and Content Characteristics using Binary Variables}
    \label{fig:Linguistic Features and Content Characteristics}
\end{figure}

\begin{figure}[!h]
    \centering
    \includegraphics[width=0.8\linewidth]{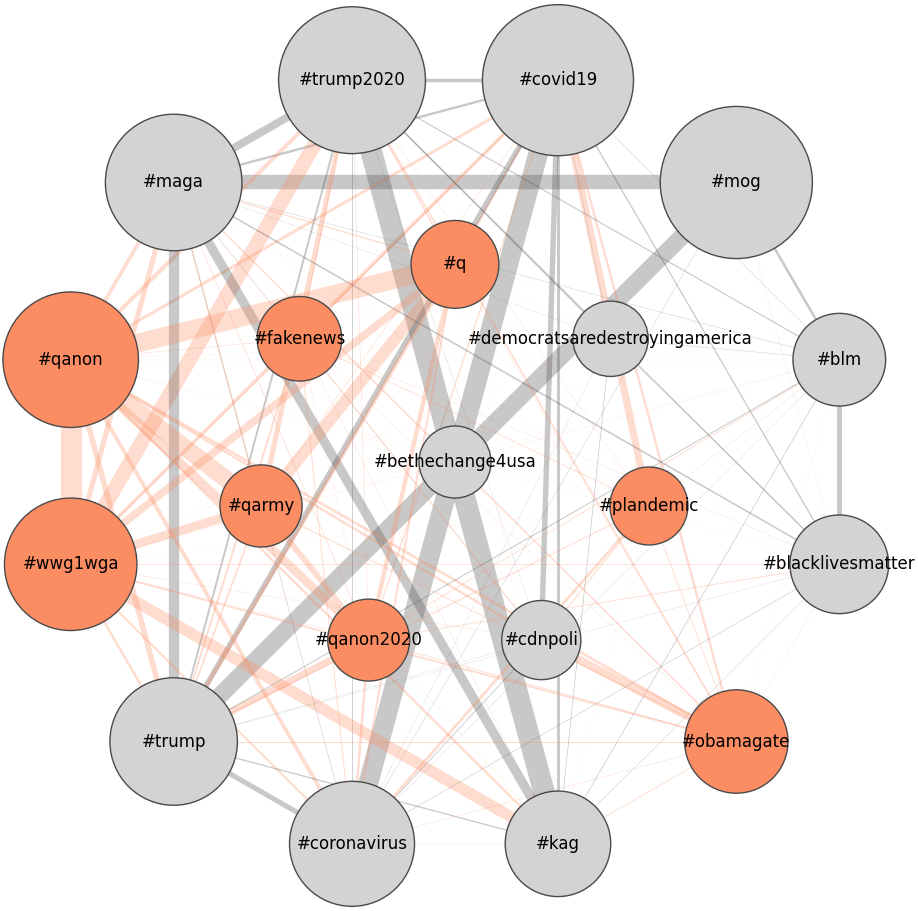}
    \caption{Visualization of Strategic Hashtag Pairing using Co-Hashtag Networks of Bot Spreaders — Node color indicates hashtag type: conspiracy-related hashtags are shown in red, non-conspiracy hashtags in gray}
    \label{fig:Co-Hashtag Networks: Bots}
\end{figure}

\begin{figure}[!h]
    \centering
    \includegraphics[width=0.8\linewidth]{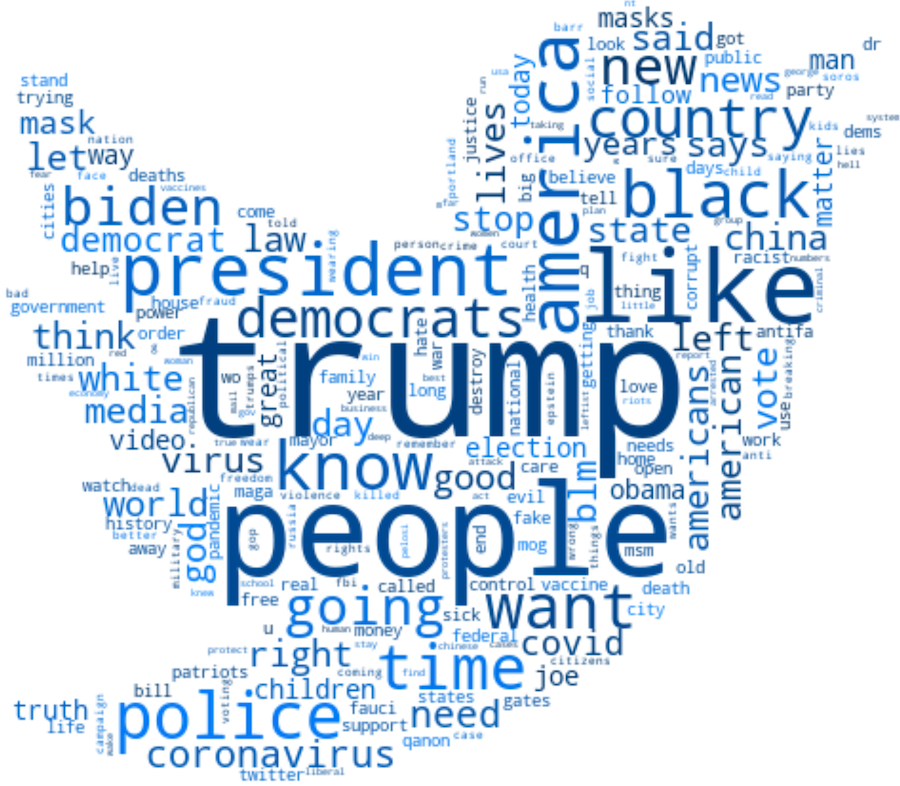}
    \caption{Visualization of the Most Frequently Used Words using Word Cloud of Bot Spreaders}
    \label{fig:Word Cloud: Bots}
\end{figure}

\begin{enumerate}
    \item \textit{Language Complexity and Readability}:
    The Flesch-Kincaid analysis reveals that Human Superspreaders have the highest mean readability score (\(M=9.02\)), approximately one grade level higher than Bot Spreaders (\(M=8.11\)) and Human Non-Spreaders (\(M=8.04\)). 
    An ANOVA confirms significant differences between groups (\(F(3, 805397) = 606.9965, p<0.0001\)). Accordingly, when crafting their tweets, Human Superspreaders appear to put a particular focus on accessibility for a broad audience.
    
    
    \item \textit{Linguistic Features and Content Characteristics}:
    Figure~\ref{fig:Linguistic Features and Content Characteristics} highlights significant differences in mentions, media, hashtags, emojis, and punctuation usage among user groups, with all differences between the groups confirmed as statistically significant by a chi-square test (\(p < 0.0001\)). Bot Spreaders use mentions and hashtags over twice as often as Human Superspreaders, include nearly three times more emojis, and utilize nearly 50\% more media items and exclamation marks. Additionally, Bot Spreaders employ all caps in 2.33\% of tweets compared to only 0.08\% for Human Superspreaders, making their usage nearly 30 times more frequent. Evidently, Bot Spreaders seem to operate under the assumption that such structural elements increase visibility.
    


    \item \textit{Hashtags and Co-occurrence Networks}:
    Table~\ref{tab:Top 10 Most Frequent Hashtags} presents the ten most frequently used hashtags across the four user groups, with hashtags common to at least three groups highlighted. Human Superspreaders primarily use hashtags like \#antifa (374 occurrences) and \#portlandriots (233 occurrences), while including \#maga (Make America Great Again) only 32 times. This aligns with their general tendency to use fewer hashtags; for example, our top Human Superspreader used only 156 hashtags across 1,440 unique tweets in the dataset. Bot Spreaders, by contrast, heavily rely on highly visible and politically charged hashtags such as \#mog (Man of God, 628 occurrences), \#trump2020 (585 occurrences), \#maga (Make America Great Again, 506 occurrences), \#qanon (498 occurrences), and \#wwg1wga (Where We Go One We Go All, 476 occurrences).
    
    Figure~\ref{fig:Co-Hashtag Networks: Bots} visualizes the strategic linkage of conspiratorial and general hashtags by Bot Spreaders. The co-hashtag graph highlights the pairing of popular hashtags like \#wwg1wga—a top hashtag in the dataset—with niche conspiratorial hashtags, including \#qarmy, \#q, \#qanon, and \#qanon2020. 

    \item \textit{Word Frequency}: 
    Table~\ref{tab:Top 10 Most Frequent Words} reveals notable differences in word usage across the four user groups. While the absolute counts differ due to variations in the number of tweets analyzed for each group, the relative order of word frequency provides valuable insights into thematic focus.

    Across all groups, the terms \textit{trump} and \textit{people} consistently ranks among the most frequently used words, highlighting the centrality of political discourse. Human Superspreaders prominently feature terms such as \textit{antifa} (\(1,067\)), \textit{democrats} (\(659\)) and  \textit{biden} (\(584\)), reflecting their alignment with political and protest-related narratives. In contrast, Bot Spreaders amplify racially sensitive topics, as evidenced by their frequent use of terms like \textit{black} (\(1,879\)) and \textit{police} (\(1,855\)). The word usage of Bot Spreaders is visually represented in a word cloud in Figure~\ref{fig:Word Cloud: Bots}.

\end{enumerate}

\textit{Sentiment, Emotion, and Toxicity Analysis.}
Similar to the previous content analysis, we now analyze a range of metrics around the sentiment, the emotion, and the toxicity of tweets between Human Superspreaders, Human Spreaders, Human Non-Spreaders, and Bot Spreaders.

\begin{enumerate}
    \item \textit{VADER Sentiment}: Figure~\ref{fig:VADER-Sentiment} shows that negative sentiment dominates all groups yet it is most pronounced among Human Superspreaders (\(58\%, M = -0.62, SD = 0.26, Var = 0.07\)). 
    An ANOVA confirms differences in mean scores for both Negative (\(F(3, 370355) = 354.90, p < 0.0001\)) and Positive sentiments (\(F(3, 322887) = 425.16, p < 0.0001\)), but not for Neutral (\(F(3, 113961) = 0.89, p = 0.4423\)). Accordingly, conspiracy-related tweets appear to circulate predominantly through content with negative sentiment.
\begin{figure}[!h]
    \centering
    \includegraphics[width=0.6\linewidth]{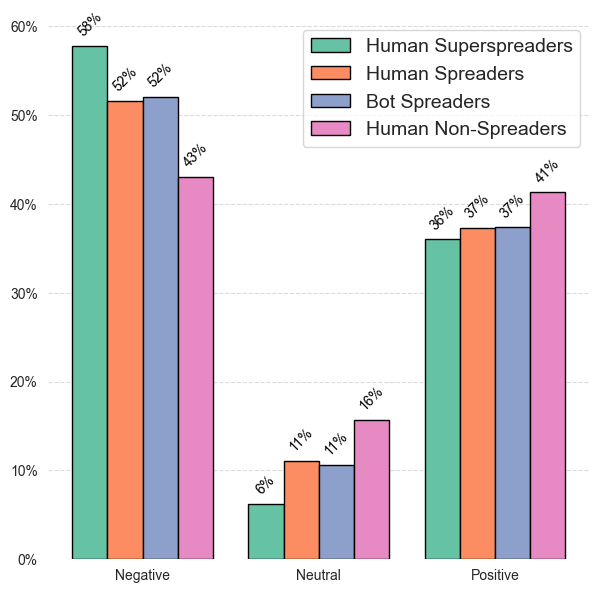}
    \caption{Analysis of Tweet Sentiment using VADER Sentiment Analysis}
    \label{fig:VADER-Sentiment}
\end{figure}

    \item \textit{Ekman’s Basic Emotions}: 
    Figure~\ref{fig:Ekman's basic emotions} reveals that Human Superspreaders and Bot Spreaders frequently use anger and fear, while Human Non-Spreaders show more neutral content.
    ANOVA confirms the differences to be statistically significant for all emotions except disgust (\(F(3, 57827) = 1.65, p = 0.176\)).

    \begin{figure}[!h]
    \centering
    \includegraphics[width=1\linewidth]{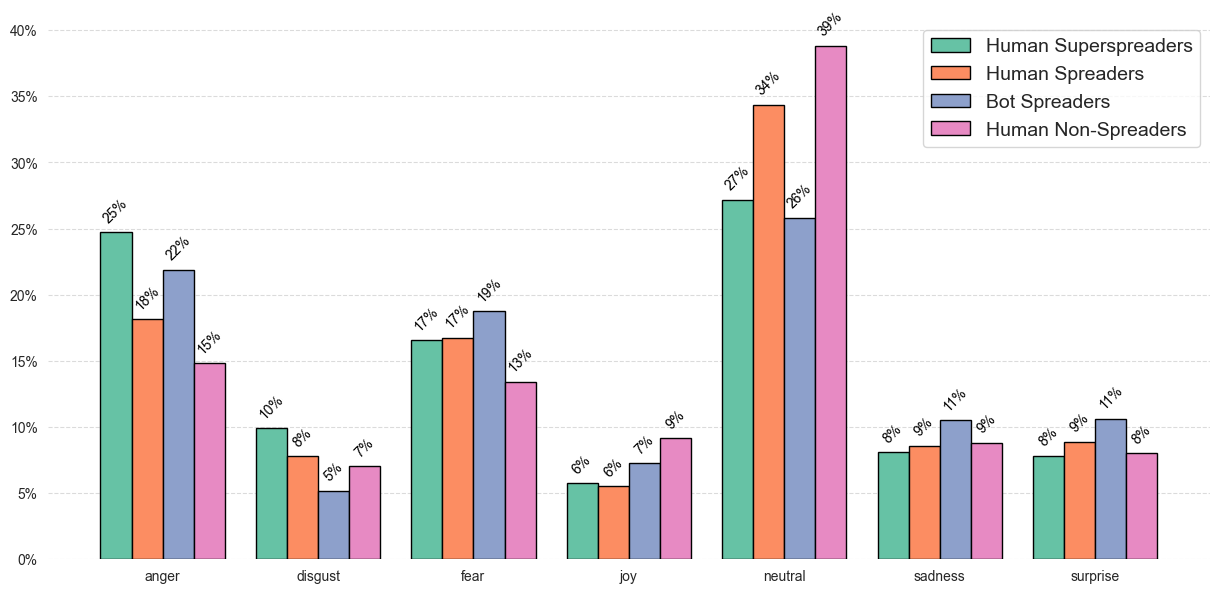}
    \caption{Analysis of Ekman’s Basic Emotions using a Fine-Tuned DistilRoBERTa Model}
    \label{fig:Ekman's basic emotions}
\end{figure}

    \item \textit{Tweet Toxicity}:
    Figure~\ref{fig:Toxicity} shows that Bot Spreaders have the highest mean toxicity scores (\(M=0.25\), \(SD=0.20\), \(Var=0.04\)), followed by Human Superspreaders (\(M=0.22\), \(SD=0.20\), \(Var=0.04\)), indicating more inflammatory content. Conversely, Human Non-Spreaders demonstrate the least toxic behavior (\(M=0.17\), \(SD=0.19\), \(Var=0.03\)). ANOVA shows the differences in the toxicity scores across the groups to be statistically significant (\(F(3, 793423) = 2958.62, p < 0.0001\)).
    
\begin{figure}[!h]
    \centering
    \includegraphics[width=0.7\linewidth]{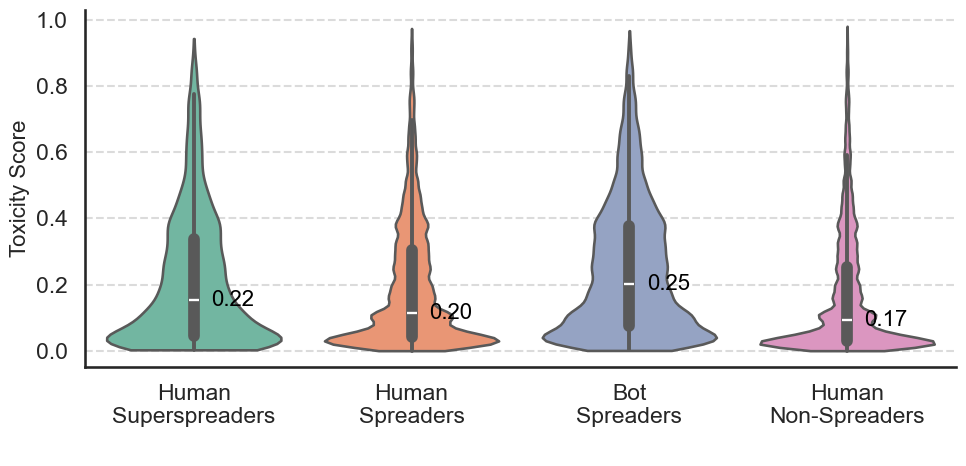}
    \caption{Analysis of Language Toxicity using Google Jigsaw Perspective API}
    \label{fig:Toxicity}
\end{figure}

\end{enumerate}

\begin{figure}[!h]
    \centering
    \includegraphics[width=1\linewidth]{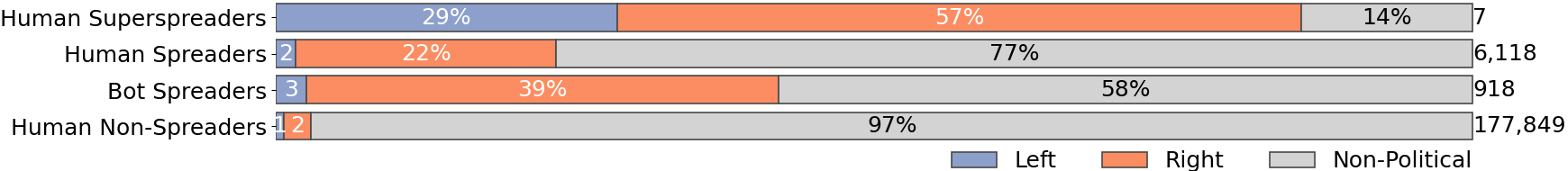}
    \caption{Analysis of Political Orientation using Hashtag-Based Classification}
    \label{fig:Political Orientation}
\end{figure}

\textit{Analysis of Political Orientation.} Using hashtag-based classification, Figure~\ref{fig:Political Orientation} shows Human Superspreaders (57\% right-leaning, 29\% left-leaning) and Bot Spreaders (39\% right-leaning, 3\% left-leaning) favor politically charged content, while Human Non-Spreaders (97\% non-political) focus on less partisan topics. Interestingly, in contrast to humans, Bots spreading conspiracy theories are predominantly right-leaning. A Chi-Square test of independence (\(\chi^2 = 11942.49, p < 0.0001, df = 6\)) indicates that the overall distribution varies significantly. 

\section{Discussion}
Previous research has shown that Human Superspreaders and Bot Spreaders play pivotal roles in amplifying harmful conspiracy content on Twitter. Our findings confirm these roles and add two contributions: effective approaches for reducing conspiracy-related content while limiting de-platforming, and a deeper understanding of how Human and Bot Spreaders differ in the nature of the content they disseminate. Understanding these differences is key to devising strategies that effectively mitigate their impact on the broader online ecosystem.

\subsection{Interpretation of Results}

\textit{Metrics to Rank and Remove Superspreaders.} The dismantling analysis reveals that the H-index and G-index stand out as particularly effective tools for identifying Human Superspreaders. Specifically, for a limited number of suspended users, they remove the largest share of conspiracy-related tweets on the platform. The H-index in particular demonstrates exceptional performance at lower levels of suspended users. We interpret this as a direct consequence of its ability to balance breadth and depth of influence, requiring users to achieve consistent engagement across multiple tweets to be classified as influential. Unlike other metrics, which may disproportionately emphasize either outlier tweets or sheer volume of activity, the H-index prioritizes accounts that sustain influence through repeated high-impact posting. In our interpretation, this characteristic is consistent with what we would expect of Human Superspreaders: rather than depending on sporadic viral spikes, their influence would be expected to accumulate through a steady cadence of tweets that regularly engage their audience. The H-index not only capitalizes on this dynamic to effectively reduce conspiracy-related content but also offers a transparent and interpretable method, aligning with calls for greater interpretability in algorithmic decision-making \citep{rudin2019stop}.


\textit{Comparison Between Human Superspreaders and Bots.}
Users that share conspiracy-related content do share some common words and hashtags (\#maga, “trump,” “president,” “people”) yet significant differences emerge in linguistic features, content characteristics, emotional expressions, toxicity, and political orientations between Human (Super)spreaders and Bot Spreaders.

First, the difference in the level of linguistic complexity between Human Superspreaders (grade 9, 14–15 years old) and Bot Spreaders (grade 8, 13–14 years old) likely reflects intentional strategies tailored to their objectives in spreading conspiracy theories. We interpret the more complex language of Human Superspreaders as signaling credibility and authority and presenting them as well-informed, whereas we read the simpler language of Bot Spreaders as emphasizing accessibility and shareability across broad audiences.

Second, Human Superspreaders tend to post longer tweets with more substantive text, relying less on structural elements such as emojis. We interpret this as maintaining a tone of seriousness and professionalism likely to signal credibility and authority. By contrast, the content characteristics of Bot Spreaders reflect a different strategy: their frequent use of hashtags and mentions enhances discoverability, allowing them to infiltrate trending conversations and target specific audiences \citep{HashtagChakrabarti}. Hashtags further enable Bot Spreaders to simulate individual identities by aligning with or opposing specific ideologies, communities, or social movements \citep{HashtagsPoliticalDirectionStewart}, while also fostering subcommunities and reinforcing bonds among like-minded users \citep{HashtagCommunityLledo}. Bots also rely more heavily on visual components such as media, which can amplify perceived intensity \citep{ExclamationMarksTeh} and convey emotions like excitement, emphasis, and urgency \citep{AllCapsHeath}. These strategies may make their messages more engaging and memorable, boosting interaction and retweet likelihood. Similarly, Bot Spreaders use emojis extensively, likely to humanize their content, enhance emotional appeal, and strengthen community ties, thereby increasing emotional resonance and audience connection \citep{EmojiTwitterQAnonWang, EmojiConspiracyGualda}.



Third, both Human Superspreaders and Bot Spreaders consistently produce high levels of negative sentiment and toxic language, which we intepret as an effort to provoke strong emotional reactions and increase engagement. This approach indeed often triggers emotions such as anger and fear, which are effective at capturing attention and motivating shares \citep{TheSpreadOfTrueAndFalseVosoughi}. 

Finally, conspiracy-spreading user groups tend to favor right-leaning content. Frequently used hashtags such as \#maga and \#trump2020, as well as references to ``trump'' or ``president'' underscore their focus on politically charged topics. Interestingly, while there is also a substantive share of Human Superspreaders that disseminate left-leaning content, Bot Spreaders almost never do. Instead, their content is overwhelmingly right-leaning. One possible interpretation is that this reflects the stronger interest of foreign actors in amplifying right-leaning misinformation, as documented by prior research \citep[e.g., in case of Russia, see][]{sharma2022characterizing}.

\subsection{Practical Implications}

Our work highlights the necessity for targeted, evidence-based interventions to mitigate the influence of Human Superspreaders and Bot Spreaders. First, the effectiveness of the H-index in identifying Human Superspreaders along with its interpretability to human stakeholders highlights its value for practitioners who may leverage the approach for automated detection and suspension of accounts. Second, for Bot Spreaders, who platforms may be more willing to suspend, the analysis uncovers pronounced differences from human users, providing a pathway for more accurate detection and effective removal. Third, by revealing the emotional and ideological tactics of Human Superspreaders versus Bot Spreaders, our findings may inform the creation of campaigns that directly challenge conspiracy narratives while promoting critical thinking and media literacy. Overall, by integrating advanced detection systems, reducing the visibility of harmful content, and critical thinking and media literacy, platforms and policymakers can build upon our findings to significantly limit the spread of conspiracy theories.

\subsection{Limitations and Future Research}

This study sheds light on the roles of Human Superspreaders and Bot Spreaders in amplifying conspiracy narratives but has limitations that may impact its generalizability. First, the dataset was limited to a specific period, purpose, and platform, which, while enabling detailed analysis, does not capture dynamics across other platforms or shifts in user behavior over time. Future research should include longitudinal data and analysis across multiple platforms such as Facebook, TikTok, and Reddit. 
Second, Our analysis was further constrained by the removal of many conspiracy-related accounts and tweets before data collection. While this enabled us to focus on actors who have thus far evaded detection, it also introduced a survivor bias, which may underestimates the influence of excluded actors. Real-time analyses or studies of banned accounts could provide a deeper understanding of network dynamics and the influence of removed actors. Third, Sentiment, emotion, and toxicity analyses relied on pre-trained models that may not fully capture cultural or linguistic nuances of online communication. Future studies could refine these models to improve accuracy, particularly for subtle emotional cues and complex toxic language. 

\section{Conclusion}
This study provides both an effective and interpretable approach to curb the spread of conspiracy theories and a comprehensive examination of the mechanisms through which Human Superspreaders and Bot Spreaders disseminate conspiracy narratives on social media. 
Specifically, an adapted H-index proved most effective in ranking Human Superspreaders according to their influence on the spread of conspiracies. A comparative analysis of Human Superspreaders, Bot Spreaders, and other user groups revealed shared traits, such as frequent political hashtags (\#antifa, \#maga) and terms ("Trump", "democrats”), emphasizing the centrality of political discourse in the propagation of conspiracy narratives. Clear differences, however, emerged in linguistic complexity, content features, emotional expressions, and political orientations, which can be leveraged to further refine detection systems or to counter conspiracy content through measures such as flagging posts and prompting users to critically reflect on them. Overall, the work provides a strong basis for mitigating conspiracy narratives, offering valuable insights for platforms, policymakers, and researchers aiming to protect online discourse in the digital age.

\section{Acknowledgments}
Some of the computing for this project was performed on the Sherlock cluster. We would like to thank Stanford University and the Stanford Research Computing Center for providing computational resources and support that contributed to these research results.

Sections of this work were generated with the assistance of ChatGPT, which was used for text formulation, grammar and spelling checks, as well as for the improvement and development of code.

\bibliography{aaai25}

\begin{table*}[!t]
\section{Appendix}

    \caption{Top Ten Most Frequently Used Hashtags}
    \centering
    \begin{subtable}{0.24\textwidth}
        \centering
        \caption{Human Superspreaders}
        \begin{tabular}{p{2.7cm}c} 
            \hline
            Hashtag & Count\\ 
            \hline\hline
            antifa & 374\\
            portlandriots & 233\\
            blacklivesmatter & 164\\
            portlandmugshots & 67\\
            endthenightmare & 40\\
            fact & 37\\
            georgefloyd & 34\\
            maga &32\\
            obamagate & 29\\
            joebiden & 18\\
        \end{tabular}
        \label{tab:SSHashtags}
    \end{subtable}
    \hfill
    \begin{subtable}{0.24\textwidth}
        \centering
        \caption{Human Spreaders}
        \begin{tabular}{p{2.5cm}c} 
            \hline
            Hashtag & Count\\ 
            \hline\hline
            covid19 &5577\\
            coronavirus & 4125\\
            qanon & 2166\\
            trump & 1749\\
            wwg1wga & 1723\\
            maga & 1463\\
            blacklivesmatter & 989\\
            covid & 947\\
            plandemic & 937\\
            kag & 903\\
        \end{tabular}
        \label{tab:NonSSHashtags}
    \end{subtable}
    \hfill
    \begin{subtable}{0.21\textwidth}
        \centering
        \caption{Bot Spreaders}
        \begin{tabular}{p{2.0cm}c} 
            \hline
            Hashtag & Count\\ 
            \hline\hline
            mog & 628\\
            covid19 & 619\\
            trump2020 & 585\\
            maga & 506\\
            qanon & 498\\
            wwg1wga & 476\\
            trump & 441\\
            coronavirus & 424\\
            kag &  302\\
            obamagate & 290\\
        \end{tabular}
        \label{tab:BotsHashtags}
    \end{subtable}
    \hfill
    \begin{subtable}{0.24\textwidth}
        \centering
        \caption{Human Non-Spreaders}
        \begin{tabular}{p{2.7cm}c} 
            \hline
            Hashtag & Count\\ 
            \hline\hline
            covid19 & 9763\\
            coronavirus & 6143\\
            blacklivesmatter & 3210\\
            trump & 2465\\
            maga & 1933\\
            foxnews & 1560\\
            blm & 1515\\
            wwg1wga & 1481\\
            trump2020 & 1446\\
            breaking & 1443\\
        \end{tabular}
        \label{tab:NonConHashtags}
    \end{subtable}
    \label{tab:Top 10 Most Frequent Hashtags}
\end{table*}

\begin{table*}[!t]
    \caption{Top Ten Most Frequently Used Words}
    \centering
    \begin{subtable}{0.22\textwidth}
        \centering
        \caption{Human Superspreaders}
        \begin{tabular}{p{1.8cm}c} 
            \hline
            Word & Count\\ 
            \hline\hline
            trump	&	2175	\\
            antifa	&	1067	\\
            people	&	1065	\\
            president	&	991	\\
            police	&	729	\\
            democrats	&	659	\\
            media	&	628	\\
            portland	&	605	\\
            biden	&	584	\\
            america	&	578	\\
        \end{tabular}
        \label{Table:NeatPulloutAB}
    \end{subtable}
    \hfill
    \begin{subtable}{0.24\textwidth}
                \centering
        \caption{Human Spreaders}
        \begin{tabular}{p{2cm}c}  
            \hline
            Word & Count\\ 
            \hline\hline
             trump	&	42321	\\
            people	&	29817	\\
            coronavirus&	14938	\\
    president& 14435	\\
            like&	14292	\\
            new	&	12579	\\
            know&	10604	\\
            time&	10283	\\
            america	&	9151	\\
            going	&	8938	\\
        \end{tabular}
        \label{Table:NeatPulloutAB}
    \end{subtable}
    \hfill
    \begin{subtable}{0.21\textwidth}
                \centering
        \caption{Bot Spreaders}
        \begin{tabular}{p{1.8cm}c}  
            \hline
            Word & Count\\ 
            \hline\hline
            trump	&	5918	\\
            people	&	4200	\\
            like	&	2479	\\
            president	&	2402	\\
            america	&	2386	\\
            know	&	1978	\\
            black	&	1879	\\
            police	&	1855	\\
            time	&	1762	\\
            want	&	1732	\\
        \end{tabular}
        \label{Table:NeatPulloutAB}
    \end{subtable}
    \hfill
    \begin{subtable}{0.23\textwidth}
                \centering
        \caption{Human Non-Spreaders}
        \begin{tabular}{p{1.8cm}c} 
            \hline
            Word & Count\\ 
            \hline\hline
            people	&	62146	\\
            trump	&	62042	\\
            like	&	35223	\\
            de	&	32791	\\
            time	&	27250	\\
            new	&	26766	\\
            know	&	24007	\\
            today	&	22858	\\
            president	&	22660	\\
            police	&	21497	\\
        \end{tabular}
        \label{Table:NeatPulloutAB}
    \end{subtable}
    \hfill
    \label{tab:Top 10 Most Frequent Words}
\end{table*}

\end{document}